\documentclass[conference,a4paper]{IEEEtran}
\IEEEoverridecommandlockouts
\usepackage[utf8]{inputenc} 
\usepackage[T1]{fontenc}
\usepackage{url}              
\usepackage{cite}             

\usepackage[cmex10]{amsmath}  
\usepackage{mleftright}       
\mleftright                   

\usepackage{graphicx}         
\usepackage{booktabs}  

\usepackage{cite}
\usepackage{amsmath,amssymb,color,graphicx,caption,subcaption,comment,tikz,url,latexsym} 
\usepackage{algorithmic}
\usepackage{graphicx}
\usepackage{textcomp}
\usepackage{pgfplots}
\usepackage{xcolor}
\usepackage{url}
\usepackage{multirow}
\usepackage{mathtools}
\usepackage[left=0.62in,right=0.62in,top=.76in,bottom=.76in]{geometry}

\usepgfplotslibrary{units}
\usetikzlibrary{spy,backgrounds}
\usepackage{pgfplotstable}

\def\BibTeX{{\rm B\kern-.05em{\sc i\kern-.025em b}\kern-.08em
		T\kern-.1667em\lower.7ex\hbox{E}\kern-.125emX}}

\usepackage{amsmath,amssymb,color,graphicx,caption,subcaption,comment,tikz,url,latexsym} 
\usepackage{graphicx}
\usepackage{textcomp}
\usepackage{pgfplots}
\usepackage{xcolor}
\usepackage{url}
\usepackage{multirow}
\usepackage{bbm}
\usepackage{dsfont}
\usepackage{mathtools}
\usepgfplotslibrary{units}
\usetikzlibrary{spy,backgrounds}
\usepackage{pgfplotstable}

\def\BibTeX{{\rm B\kern-.05em{\sc i\kern-.025em b}\kern-.08em
		T\kern-.1667em\lower.7ex\hbox{E}\kern-.125emX}}

\newtheorem{theorem}{Theorem} 

\newtheorem{definition}{Definition}
\newtheorem{lemma}{Lemma}

\newtheorem{corollary}[theorem]{Corollary}

\usepackage{bbm}

\newcommand{\E}{\mathbb{E}}

\begin{document}	
	\interdisplaylinepenalty=0
	\title{Strong  Converses for  Memoryless Bi-Static ISAC}

	\author{
		\IEEEauthorblockN{Mehrasa Ahmadipour\IEEEauthorrefmark{1}, Mich\`ele Wigger\IEEEauthorrefmark{2}, Shlomo Shamai\IEEEauthorrefmark{3} 
		}
	\IEEEauthorblockA{\small\IEEEauthorrefmark{1}
		UMPA, ENS de Lyon
	\url{mehrasa.ahmadipour@ens-lyon.fr}}
		\IEEEauthorblockA{\small\IEEEauthorrefmark{2} LTCI Telecom Paris, IP Paris, 91120 Palaiseau, France, Emails:
			\url{michele.wigger@telecom-paris.fr}}
		\IEEEauthorblockA{\small\IEEEauthorrefmark{3} Technion, Israel ,  Email: sshlomo@ee.technion.ac.il
		}
	}

	
	\allowdisplaybreaks[4]
	\sloppy
	\maketitle
	\begin{abstract}
The paper characterizes  the fundamental limits of  integrated sensing and communication (ISAC) systems with a bi-static radar, where the  radar receiver is located close to the transmitter  and  estimates or detects the state based on the transmitter's channel inputs and the backscattered signals. Two models are considered. In the first model, the memoryless state sequence is distributed according to a fixed distribution and the goal of the radar receiver is to reconstruct this state-sequence with  smallest possible distortion. In the second model, the memoryless state is distributed either according to $P_S$ or to $Q_S$ and the radar's goal is to detect this underlying distribution so that the missed-detection error probability has maximum exponential decay-rate  (maximum Stein exponent). Similarly to previous results, our fundamental limits show that the tradeoff between  sensing and communication  solely stems from the empirical statistics of the transmitted codewords which influences both  performances. The main technical  contribution are two strong converse proofs that hold for all probabilities of communication error $\epsilon$ and excess-distortion probability or false-alarm probability $\delta$ summing to less than 1, $\epsilon+\delta < 1$. These proofs are based on  two parallel change-of-measure arguments on the sets of typical sequences, one change-of-measure to obtain the desired bound on the communication rate, and the second to bound the sensing performance.
\end{abstract}
\section{Introduction}
Sensing is a promising new feature in the upcoming 6G  \cite{Tong} and  WIFI standards \cite{Wifi}. In  particular,  huge technological efforts are  being made to integrate radar systems with communication systems. Such integrated systems are especially appealing for autonomous driving applications or autonomous manufacturing sites (as part of the Industry 4.0). In \emph{integrated sensing and communication systems (ISAC)}, the idea is to use the backscattered signals from communication for radar applications  to sense the environment, detect hazardous events, or infer properties of other terminals (e.g., velocities or directions of other cars). %

\begin{figure}[t!]
\includegraphics[width=9cm]{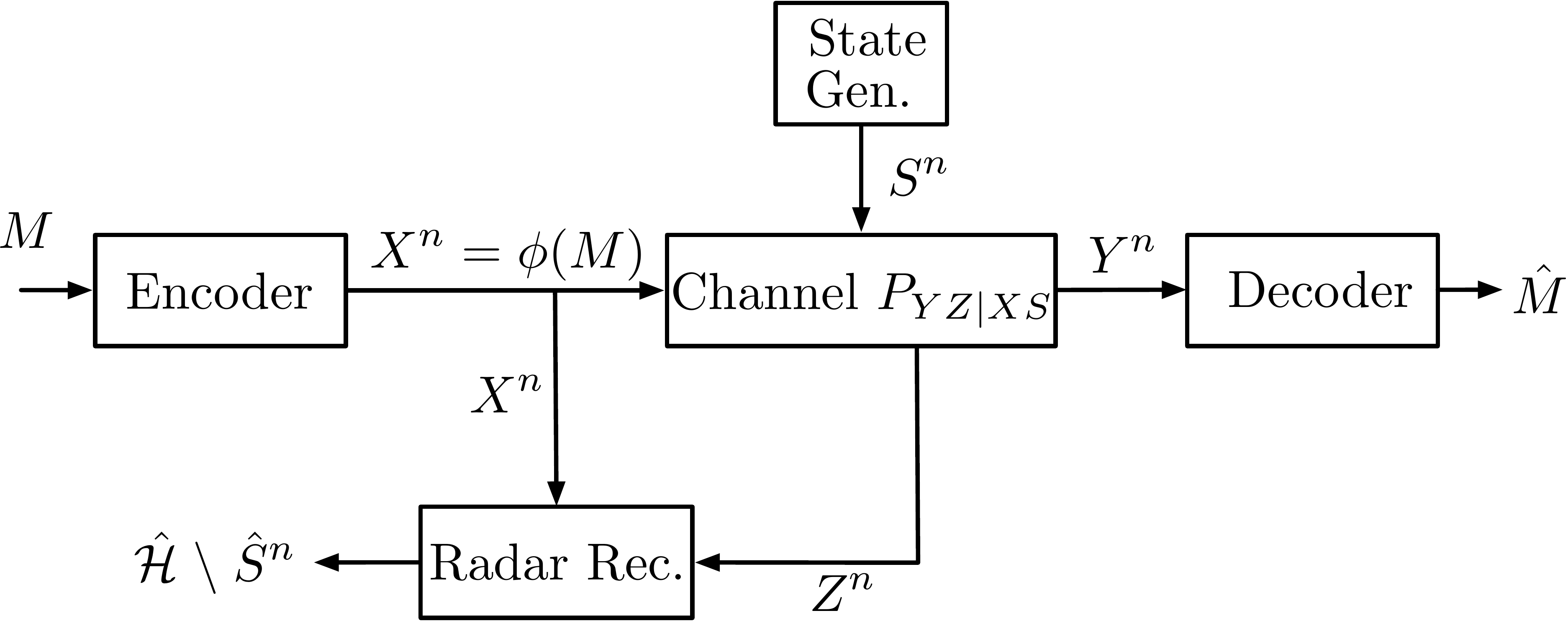}
\caption{Bistatic Radar ISAC Model}
\label{fig:channel}
\end{figure}

	While ISAC has inspired  a  plethora of works in the signal processing and communications community, see  for example \cite{dokhanchi2019adaptive,ISAC_Liu,ISAC_OFDM,ISAC_radar,ISAC_radar2,LiuSurvey2022} and references therein, only few works were reported from the  information-theoretic community \cite{kobayashi2018joint,kobayashi2019joint, Mehrasa2022IT,  Mehrasa2022MACISIT, Gunlu,Joudeh2021JBinaryDetect,Joudeh2022Discrim, Bloch2022}. The results in \cite{Joudeh2021JBinaryDetect, Bloch2022,Joudeh2022Discrim} and the present manuscript all  focus on the system model in Figure~\ref{fig:channel} consisting of a transmitter (Tx) sending a message to a receiver (Rx) over a state-dependent discrete memoryless channel (SDMC). A bistatic radar close to the Tx receives the backscattered signal modeled through generalized feedback. Due to the proximity to the Tx, this radar receiver also knows the Tx's  channel inputs and compares them to its feedback outputs.

	The works in \cite{Joudeh2021JBinaryDetect, Bloch2022 ,Joudeh2022Discrim} determined the fundamental performance limits of a  detection-version of the model in Figure~\ref{fig:channel}.  Specifically in these works, the state sequence $\{S_t\}$ is assumed constant over time, taking on one of multiple possible values depending on a underlying hypothesis, and the radar receiver aims to guess this hypothesis. Sensing performance is measured in terms of exponential decay-rate of the probability of error, either the minimum exponential decay-rate over all hypotheses  \cite{Joudeh2021JBinaryDetect, Bloch2022}  or the set of decay-rates that are simultaneously achievable under the different hypotheses \cite{Joudeh2022Discrim}. The work \cite{Bloch2022} also studies a  mono-static version of this problem, assuming that Tx coincides with the  radar receiver and thus can use the generalized-feedback signals also for communication purposes. For this mono-static radar scenario however only a coding scheme but no converse is presented. The problem is known to be hard as it relates to the challenging close-loop controlled sensing problem \cite{controlled_sensing}. 
	
	The first information-theoretic work  \cite{kobayashi2018joint} on ISAC determined the fundamental limits of the  rate-distortion version of the ISAC problem in Figure~\ref{fig:channel}. assuming that the Tx can use the feedback signals also for coding (i.e., under the close-loop coding assumption).   Extensions to network scenarios and  to scenarios with secrecy constraints were subsequently presented in \cite{kobayashi2019joint, Mehrasa2022IT,  Mehrasa2022MACISIT, Gunlu}.

In this paper, we consider both the rate-distortion version and the hypothesis testing  versions of the model in Figure~\ref{fig:channel}. In our first model, the state sequence $\{S_t\}_{t\geq 1}$ is independent and identically  distributed (i.i.d.) according to a given distribution $P_S$ and the radar wishes to  reconstruct this state sequence with smallest possible distortion. In our second model, the state-sequence $\{S_t\}$ depends on a binary hypothesis $\mathcal{H}\in\{0,1\}$. If $\mathcal{H}=0$, then $\{S_t\}$ is i.i.d. according to a distribution $P_S$ or if $\mathcal{H}=1$, it is i.i.d.  according to a distribution $Q_S$. We measure sensing performance in terms of Stein's exponent, i.e., in terms of  the maximum exponential decay-rate of the missed-detection error probability (detecting $P_S$  instead of $Q_S$) under a permissible threshold on the false-alarm probability (detecting $Q_S$  instead of $P_S$). 
	
	For both our models, we determine the fundamental limits of  communication rates and  distortion/missed-detection error exponents that are simultaneously achievable. 
	Similarly to previous works \cite{kobayashi2018joint,Joudeh2021JBinaryDetect, Bloch2022}  our  limits  exhibit a tradeoff between the  sensing and communication performances, which however solely stems from the empirical statistics of the codewords used for communication.

	The direct parts of our proofs follow immediately from existing works. Our contributions are the proofs of the converse results. In fact, we present strong converse proofs  that  hold whenever the maximum allowed probability of communication error $\epsilon$ and the maximum allowed distortion-excess probability or false-alarm probability $\delta$ satisfy $\epsilon+\delta < 1$. The converse proofs are extensions of the channel coding strong converse proof in \cite{strongIT} to incorporate also the sensing bounds. Interestingly, the same change-of-measure as in \cite{strongIT} can be used to obtain the desired bound on the rate of communication. Different  changes-of-measure are used to obtain the desired bounds on the sensing performances.
	
	Strong converse proofs based on change-of-measure arguments go back to  Gu and Effros \cite{GuEffros_1,GuEffros_2} and can be also found in various other works, e.g., \cite{HanKobayashi}. The proof method was formalized and first applied to channel coding by Tyagi and Watanabe \cite{tyagi2019strong}. Recent works \cite{strongIT, mustapha, sara} slightly modified and simplified the technique in \cite{tyagi2019strong} by restricting the new measures to sequences on typical or conditionally-typical sets. This feature allows to circumvent resorting to variational characterizations for the multi-letter and single-letter problems as proposed in \cite{tyagi2019strong}. Notice that the works \cite{mustapha,sara} also showed the utility of the proposed converse proof method for scenarios with expectation constraints, in which case the fundamental limits depend on the permissible error probabilities.


\textit{Notation:}
Upper-case letters are used for random quantities and lower-case letters for deterministic realizations. Calligraphic font is used for sets.   All random variables are assumed finite and discrete. We  abbreviate   the $n$-tuples $(X_1,\ldots, X_n)$ and $(x_1,\ldots, x_n)$ as $X^n$ and $x^n$ and the $n-t$ tuples  $(X_{t+1},\ldots, X_n)$ and $(x_{t+1},\ldots, x_n)$ as $X_{t+1}^n$ and $x_{t+1}^n$. We further abbreviate \emph{independent and identically distributed} as \emph{i.i.d.} and \emph{probability mass function} as \emph{pmf}.

  Entropy, conditional entropy, and mutual information functionals are written as $H(\cdot)$, $H(\cdot|\cdot)$, and $I(\cdot;\cdot)$, where the arguments of these functionals are random variables and whenever their probability mass function (pmf)  is not clear from the context, we add it as a subscript  to these functionals. The Kullback-Leibler divergence between two pmfs is denoted by  $D(\cdot \| \cdot)$.  We shall use  $\mathcal{T}_{\mu}^{(n)}(P_{XY})$  to indicate the jointly strongly-typical set with respect to the pmf $P_{XY}$ on the product alphabet $\mathcal{X}\times \mathcal{Y}$ and parameter $\mu$ as defined in  \cite[Definition 2.8]{Csiszarbook}. Specifically, denoting by $n_{x^n,y^n}(a,b)$ the number of occurrences of the pair $(a,b)$ in sequences $(x^n,y^n)$: 
  \begin{equation}
  n_{x^n,y^n}(a,b) =  \left| \{t\colon (x_t,y_t)=(a,b)  \}\right| ,
  \end{equation}
 a pair  $(x^n,y^n)$ lies in $ \mathcal{T}_{\mu}^{(n)}(P_{XY})$ if 
\begin{equation} \left|  \frac{n_{x^n,y^n}(a,b)}{n}   - P_{XY}(a,b) \right| \leq \mu,  \qquad \forall (a,b)\in\mathcal{X}\times \mathcal{{Y}}, 
\end{equation}
and $n_{x^n,y^n}(a,b)=0$ whenever  $P_{XY}(a,b)=0$.
   The conditionally strongly-typical set with respect to a conditional pmf $P_{Y|X}$ from $\mathcal{X}$ to $\mathcal{Y}$,  parameter $\mu>0$, and  sequence $x^n\in \mathcal{X}^n$ is denoted $\mathcal{T}_{\mu}^{(n)}(P_{Y|X}, x^n)$ \cite[Definition 2.9]{Csiszarbook}. It contains all sequences $y^n\in\mathcal{Y}^n$ satisfying 
 \begin{equation} \left|  \frac{n_{x^n,y^n}(a,b)}{n}   -  \frac{n_{x^n}(a)}{n} P_{Y|X}(b|a) \right| \leq \mu,  \qquad \forall (a,b)\in\mathcal{X}\times \mathcal{{Y}}, 
\end{equation}
and  $n_{x^n,y^n}(a,b)=0$ whenever $P_{Y|X}(b|a)=0$.  Here $n_{x^n}(a)$ denotes the number of occurrences of symbol $a$ in $x^n$. In this paper, we denote the joint type of $(x^n,y^n)$ by $\pi_{x^ny^n}$, i.e., 
\begin{equation}
\pi_{x^ny^n}(a,b)\triangleq\frac{n_{x^n,y^n}(a,b)}{n}.
\end{equation}
 Accordingly, the marginal type of $x^n$ is written as $\pi_{x^n}$.



\section{Memoryless State and Average Distortion as a Sensing Measure}\label{sec:distortion}

Consider the bistatic radar receiver model over a memoryless  channel in Fig.~\ref{fig:channel}. A transmitter (Tx) that wishes to communicate a random message $M$ to a receiver (Rx) over a state-dependent channel. The message $M$ is uniformly distributed over the set $\{1,\ldots, 2^{nR}\}$ with $R>0$ and $n>0$  denoting the  rate and  blocklength of communication, respectively. The channel from the Tx to the Rx depends on a  state-sequence $S^n=(S_1,\ldots, S_n)$ which is i.i.d. according to a given pmf $P_S$.

For a given blocklength $n$, the Tx thus produces the $n$-length sequence of  channel inputs  
\begin{equation}
X^n = \phi^{(n)} (M)
\end{equation}
for some choice of the   encoding function $\phi^{(n)}\colon  \{1,\ldots, 2^{nR}\} \to \mathcal{X}^n$.

Based on $X^n$ and $S^n$ the channel produces the sequences $Y^n$ observed at the Rx and the backscattered signal $Z^n$. The channel is assumed memoryless and described by the stationary  transition law $P_{YZ|XS}$ implying that the pair $(Y_t,Z_t)$ is produced according to the channel law $P_{YZ|XS}$ based on the time-$t$ symbols $(X_t,S_t)$.

The Rx attempts to guess message $M$ based on the sequence of channel outputs $Y^n$:
\begin{equation}
\hat{M} = g^{(n)}(Y^n)
\end{equation}
using a decoding function of the form $g^{(n)} \colon \mathcal{Y}^n \to \{1,\ldots, 2^{nR}\}$. 

Performance of  communication  is measured in terms of average error probability 
\begin{equation}\label{eq:maxerror}
p^{(n)}(\textnormal{error}) :=  \Pr[ \hat{M} \neq M ]
\end{equation}

The radar receiver produces as a guess a reconstruction of the state sequence 
\begin{equation}
\hat{S}^n = h^{(n)}(X^n,Z^n),
\end{equation}
based on the inputs and backscattered signals. Radar sensing performance is measured as average expected distortion
\begin{equation}\label{eq:exp_disto}
\textnormal{dist}^{(n)}\left( \hat{S}^n , S^n\right)\triangleq\frac{1}{n} \sum_{i=1}^n   d \left(\hat{S}_i, S_i\right),
\end{equation}
for a given and bounded distortion function $d(\cdot,\cdot)$. 
\medskip

In this context we have the following definition and result. 

\begin{definition}
A rate-distortion pair $(R,D)$ is  $(\epsilon,\delta)$-achievable over the state-dependent  channel $(\mathcal{X},\mathcal{Y}, P_{Y|XS})$ with state-distribution $P_S$, if there exists a sequence of encoding, decoding, and estimation functions $\{(\phi^{(n)}, g^{(n)}, h^{(n)})\}$ such that  the average  probability of error satisfies
\begin{equation}\label{eq:error_eps}
\varlimsup_{n\to \infty}  p^{(n)}(\textnormal{error}) \leq  \epsilon
\end{equation}
and the excess distortion probability
\begin{equation}\label{eq:Dist}
\varlimsup_{n\to \infty} \Pr\left[ \textnormal{dist}^{(n)}\left( \hat{S}^n, S^n\right) > D\right]\leq \delta.
\end{equation}
\end{definition}

\begin{theorem}
For any $\epsilon+\delta <1 $, a rate-distortion pair $(R,D)$ is $(\epsilon,\delta)$-achievable, if and only if, there exists a pmf $P_X$ 
 satisfying
 \begin{equation}
R = I_{P_{X}P_SP_{Y|XS}}(X;Y) 
\end{equation}
and 
 \begin{equation}
D \geq  \mathbb{E}_{P_{X}P_SP_{Z|XS}}\left[  d(\hat{s}(X, Z), \ S) \right] 
\end{equation}
where 
\begin{equation}
\hat{s}(x,z) :=\min_{\hat{s}\in\hat{\mathcal{S}}}  \sum_{s} P_{S|XZ}(s|x,z) d( \hat{s}, s)
\end{equation}
and \begin{equation}
P_{S|XZ}(s|x,z) := \frac{P_S(s) P_{Z|XS}(z|x,s)}{\sum_{s'} P_{S}(s') P_{Z|XS}(z|x,s')}
\end{equation}

\end{theorem} 

\begin{IEEEproof}
The limiting case $\epsilon, \delta \downarrow 0$ of the theorem was already proved in \cite{Mehrasa2022IT}. Achievability of the theorem  follows thus directly from  this previous result. The converse is proved in the following 
subsection, also using the next lemma, which is from \cite{Mehrasa2022IT}.\end{IEEEproof}

\begin{lemma}[From \cite{Mehrasa2022IT}]\label{lem:optimal_estimator}
Without loss in optimality, one can restrict to the per-symbol estimator
\begin{equation}\label{eq:symbol}
h^{(n)}(x^n,z^n)= \left(\hat{s}(x_1,z_1), \ldots, \hat{s}(x_n,z_n) \right). 
\end{equation}
\end{lemma}

	\subsection{Strong Converse Proof}\label{sec:converseD}

Fix a sequence of  encoding and decoding functions $\{(\phi^{(n)}, g^{(n)})\}_{n=1}^\infty$ and consider the  optimal estimator $h^{(n)}$ in Lemma~\ref{lem:optimal_estimator}. Assume that   \eqref{eq:error_eps}  and \eqref{eq:Dist} are satisfied. For readability, we will also write $x^n(\cdot)$ for the function $\phi^{(n)}(\cdot)$. 
Choose a sequence of small positive numbers  $\{\mu_n \}_{n=1}^\infty$ satisfying
\begin{eqnarray}
	\lim_{n\to \infty} \mu_n  & =& 0 \label{eq:mun1d}\\
	\lim_{n\to \infty} \left(n \cdot \mu_n^2  \right)^{-1}& =& 0. \label{eq:mun2d}
\end{eqnarray} 

\medskip 

\textbf{Expurgation:}
Fix $\eta \in (0,1-\epsilon-\delta]$ and let  $\tilde{\mathcal{M}}$ be the set of messages $m$ that satisfy the following two conditions:
\begin{subequations}\label{eq:sub_exp}
	\begin{IEEEeqnarray}{rCl} 
		\Pr\left[\hat{M} \neq M|M=m \right] &\leq&  1-\eta \IEEEeqnarraynumspace \label{eq:sub_exp1} \\
		\Pr\left[ \textnormal{dist}^{(n)}\left( \hat{S}^n, S^n\right) > D |M=m \right] &\leq& 1-\eta.\label{eq:sub_exp2} 
	\end{IEEEeqnarray}
\end{subequations}
Since the set of messages  not satisfying  \eqref{eq:sub_exp1} is at most of size 
\begin{equation}\label{eq:bound_size}
	\frac{\epsilon}{(1-\eta)} 2^{nR},
\end{equation}
and similarly also the set of 
messages  not satisfying  \eqref{eq:sub_exp2} is of size at most  $\frac{\delta}{(1-\eta)}2^{nR}$, we can deduce that the set $\tilde{\mathcal{M}}$ (which is the complement of the union of these two sets) is of size at least 
\begin{equation}
	\left(1-  \frac{\epsilon+\delta}{1-\eta}\right)2^{nR}=\frac{(1-\eta - \epsilon-\delta)}{(1-\eta)}2^{nR}.\end{equation}
Define the random variable $\tilde{M}$ to be uniform over the set $\tilde{\mathcal{M}}$ and let 
\begin{equation}\tilde{X}^n=x^n(\tilde{M}),
\end{equation}
and thus 
\begin{equation}
	\frac{|\tilde{\mathcal{M}}|}{2^{nR}} \geq \left(1-  \frac{\epsilon+\delta}{1-\eta}\right)= : \gamma.\end{equation}

Let $T$  be a uniform random variable over $\{1,\ldots, n\}$, independent of all other random variables and notice that 
\begin{IEEEeqnarray}{rCl} 
	P_{\tilde{X}_T}(x) &=& \frac{1}{n} \sum_{t=1}^n P_{\tilde{X}_t(x)} \\
	& = & \frac{1}{n} \sum_{t=1}^n \mathbb{E}[\mathbbm{1}\{ \tilde{X}_t(\tilde{M})=x\}]\\
	&= &  \mathbb{E}[\pi_{{x}^{n}(\tilde{M})}(x) ]. 
\end{IEEEeqnarray}
Let now $\{n_i\}$ be an increasing subsequence of blocklengths so that the probability vector $P_{\tilde{X}_T}$ converges and denote the convergence point by $P_X$:
\begin{IEEEeqnarray}{rCl}\label{eq:convergence_point}
	\lim_{n_i \to \infty}  \frac{1}{|\tilde{\mathcal{M}}| }\sum_{m\in \tilde{M}}    \pi_{{x}^{n_i}(m)}(x) 
& =: & P_X(x), 
	\qquad \forall x \in \mathcal{X}. \IEEEeqnarraynumspace
\end{IEEEeqnarray} 
In the remainder of this proof, we restrict attention to this subsequence of blocklengths $\{n_i\}$.

\medskip

\textbf{Proof of Channel Coding Bound:}
We first prove the converse bound for channel coding. To this end, consider the two conditions
\begin{subequations}\label{eq:sub}
	\begin{IEEEeqnarray}{rCl} 
		g^{(n)}\left( y^n\right) &= &m  \nonumber \\\label{eq:cond1}\\
		\left| \pi_{s^n,x^n(m),y^n}(a,b,c) - P_S(a) \pi_{x^n(m)}(b) P_{Y|XS}(c|a,b) \right| &\leq & \mu_n \label{eq:cond3}, \nonumber\\
	\end{IEEEeqnarray}
\end{subequations}
and define for  each message $m \in \tilde{\mathcal{M}}$ the set 
\begin{IEEEeqnarray}{rCl} 
	\mathcal{D}_{\mathcal{C},m}& :=&  \left\{  (s^n,y^n)  \colon   \quad  \eqref{eq:cond1} \; \textnormal{and} \; \eqref{eq:cond3} \right\}. \IEEEeqnarraynumspace 
\end{IEEEeqnarray}
Introduce the   new random variables $({S}^n_{\mathcal{C}},{Y}^n_{\mathcal{C}})$ of  joint conditional pmf
\begin{IEEEeqnarray}{rCl} 
	\lefteqn{
		P_{S^n_{\mathcal{C}} {Y}^{n}_{\mathcal{C}}|\tilde{M}} (s^n,y^n|m) } \nonumber \\
	&=&  \frac{ P_S^{\otimes n}(s^n) \cdot P_{Y|XS}^{\otimes n}(y^n |x^n(m), s^n)}{  \Delta_{\mathcal{C},m}} \cdot \mathbbm{1} \left\{ (s^n,y^n) \in  \mathcal{D}_{\mathcal{C},m} \right\},\nonumber\\
\end{IEEEeqnarray}
for 
\begin{IEEEeqnarray}{rCl}\label{eq:Delta}
	\lefteqn{\Delta_{\mathcal{C},m} :=  \sum_{s^n,y^n} P_S^{\otimes n}(s^n) \cdot P_{Y|XS}^{\otimes n}(y^n |x^n(m), s^n)  } \hspace{4cm} \nonumber\\
	& & \cdot   \mathbbm{1} \left\{ (s^n,y^n) \in  \mathcal{D}_{\mathcal{C},m} \right\}.\IEEEeqnarraynumspace
\end{IEEEeqnarray}

By  \cite[Remark to Lemma~2.12]{Csiszarbook} and Conditions \eqref{eq:sub_exp1} and \eqref{eq:sub}, we have:
\begin{IEEEeqnarray}{rCl}\label{eq:conditions}
	\Delta_{\mathcal{C},m}\geq \eta-\frac{|\mathcal{S}||\mathcal{X}||\mathcal{Y}|}{4\mu_n^2 n}, \quad \forall m \in \tilde{\mathcal{M}}.
\end{IEEEeqnarray} 
Moreover, for 
$\tilde{M}=m$:
\begin{IEEEeqnarray}{rCl}\lefteqn{
		P_{{Y}^{n}_{\mathcal{C}}|\tilde{M}=m}(y^{n}) }\nonumber \\
	&= &  \sum_{s^n} \frac{ P_S^{\otimes n}(s^n) \cdot P_{Y|XS}^{\otimes n}(y^n |x^n(m), s^n)}{  \Delta_{\mathcal{C},m}} \cdot \mathbbm{1} \left\{ (s^n,y^n) \in  \mathcal{D}_{\mathcal{C},m} \right\} \nonumber \\\\
	&\leq &\sum_{s^n} \frac{ P_S^{\otimes n}(s^n) \cdot P_{Y|XS}^{\otimes n}(y^n |x^n(m), s^n)}{  \Delta_{\mathcal{C},m}} \\
	& = & \frac{ P_{Y|X}^{\otimes {n}}(y^{n}|x^{n}(m))}{\Delta_{\mathcal{C},m}}. \label{eq:upperPY}
\end{IEEEeqnarray}


Continue to notice that:
\begin{IEEEeqnarray}{rCl}
	R\hspace{-0.05cm}& = &\hspace{-0.05cm}\frac{1}{n} H(\tilde{M})    - \frac{1}{n} \log  \gamma   \\
	& \stackrel{(a)}{=} &\frac{1}{n}  I( \tilde{M} ; {Y}_{\mathcal{C}}^n)-  \frac{1}{n} \log  \gamma \\
	& =&\frac{1}{n}  H({Y}^n_{\mathcal{C}}) -\frac{1}{n} H\left({Y}_{\mathcal{C}}^n \big |\tilde{M} \right)- \frac{1}{n} \log  \gamma  \\
	& 	\leq 	&  \frac{1}{n}  \sum_{i=1}^n H({Y}_{\mathcal{C},i}) 	-	\frac{1}{n} H\left({Y}_{\mathcal{C}}^n\big |\tilde{M}\right) - \frac{1}{n} \log\gamma    \\
	&= &H\left({Y}_{\mathcal{C},T} \Big |T\right)- \frac{1}{n} H\left(Y_{\mathcal{C}}^n\big|\tilde{M}\right) - \frac{1}{n} \log\gamma   \\
	&\leq & H\left({Y}_{\mathcal{C},T}\right)- \frac{1}{n} H\left({Y}_{\mathcal{C}}^n\big|\tilde{M}\right)   -\frac{1}{n} \log \gamma ,\label{eq:diff}
\end{IEEEeqnarray}
where we defined the random variable $T$ to be uniform over $\{1,\ldots,n\}$ independent of the other random variables.  Here, $(a)$ holds because $\tilde{M}=g({Y}^n_{\mathcal{C}})$ by Condition~\eqref{eq:cond1}.

%

Notice next that 
\begin{IEEEeqnarray}{rCl}
	\lefteqn{P_{\tilde{X}_{T}{S}_{\mathcal{C,T}{Y}_{\mathcal{C},T}}(x,s,y)}} \hspace{.3cm}  \\
	&=&\frac{1}{n}  \sum_{t=1}^n P_{\tilde{X}_{t}{S}_{\mathcal{C},t}{Y}_{\mathcal{C},t}}(x,s,y)\label{eq:one} \\
	&=&\frac{1}{n}  \sum_{t=1}^n \mathbb{E}\left[ \mathbbm{1}\left\{\tilde{X}_{t},{S}_{\mathcal{C},t},{Y}_{\mathcal{C},t})=(x,s,y)\right\}\right]\label{eq:one} \\
	&= &\E\left[  \pi_{{x}^n(\tilde{M})S_{\mathcal{C}}^n{Y}_{\mathcal{C}}^n}(x,s,y)\right]  
\end{IEEEeqnarray}
However, by Condition~\eqref{eq:cond3} 
for any triple $(x,s,y)$ with positive $P_{S}(s) P_{Y|XS}(y|x,s)$ the following inequality is satisfied with probability 1:
\begin{IEEEeqnarray}{rCl}\label{eq:intermediate}
	\lefteqn{\left|\pi_{{x}^n(m)S^n_{\mathcal{C}}{Y}_{\mathcal{C}}^n}(x,s,y) -  \pi_{{x}^n(m)}(x) P_{Y|XS}(y|x,s)P_S(s) \right|} \nonumber \\
	 && \hspace{7cm}\leq \mu_n. \IEEEeqnarraynumspace
\end{IEEEeqnarray}


Notice that  by \eqref{eq:convergence_point} and \eqref{eq:intermediate} and 
since $\mu_{n_i}\to 0$ as $n_i\to \infty$:\begin{IEEEeqnarray}{rCl}
	\lim_{{i} \to \infty} P_{ \tilde{X}_T{S}_{\mathcal{C},T}{Y}_{\mathcal{C},T}}(x,s,y)= {P}_X(x) P_{S}(s)P_{Y|XS}(y|x,s), \label{eq:PYlimit} \IEEEeqnarraynumspace
\end{IEEEeqnarray}
which by continuity of the entropy functional implies 
\begin{equation}\label{eq:limitHy}
	\lim_{n_i\to \infty}  H\left({Y}_{\mathcal{C},T}\right) =H_{P_XP_SP_{Y|XS}}(Y).
\end{equation}

Next, by  definition and by \eqref{eq:upperPY}:
\begin{IEEEeqnarray}{rCl}
	\lefteqn{\frac{1}{{n_i}} H({Y}^{n_i}_{\mathcal{C}} |\tilde{M}=m) } \nonumber \\
	& = &- \frac{1}{{n_i}}  \sum_{y^n \in \mathcal{D}_{\mathcal{C},m}} P_{{Y}^{n_i}_{\mathcal{C}}|\tilde{M}=m}(y^{n_i}) \log P_{{Y}_{\mathcal{C}}^{n_i}|\tilde{M}=m}(y^{n_i}) \IEEEeqnarraynumspace\\
	& \geq  &- \frac{1}{n_i}  \sum_{y^{n_i} \in \mathcal{D}_{\mathcal{C},m}} P_{{Y}^{n_i}_{\mathcal{C}}|\tilde{M}=m}(y^{n_i}) \log \frac{P_{Y|X}^{\otimes n}(y^{n_i}|x^{n_i}(m))}{ \Delta_{\mathcal{C},m}} \nonumber \\ \\
	& =& -\frac{1}{n_i} \sum_{t=1}^{n_i}  \sum_{y^{n_i} \in \mathcal{D}_{\mathcal{C},m}} P_{{Y}^{n_i}_{\mathcal{C}}|\tilde{M}=m}(y^{n_i}) \log P_{Y|X}(y_t|x_t(m)) \nonumber\\
	&&  +\frac{ 1}{n_i} \log \Delta_{\mathcal{C},m}, \\
	& =& -\frac{1}{n_i} \sum_{t=1}^{n_i}  \sum_{y_t\in\mathcal{Y}} P_{{Y}_{\mathcal{C},t}|\tilde{M}=m}(y_t) \log P_{Y|X}(y_t|x_t(m)) \nonumber \\
	&& +\frac{ 1}{n_i} \log \Delta_{\mathcal{C},m}, \\
	& =& -\frac{1}{n_i} \sum_{t=1}^{n_i}  \sum_{y\in\mathcal{Y}}  \E\left[   \mathbbm{1}\left\{ {Y}_{\mathcal{C},t}=y \right \} \Big|\tilde{M}=m\right] \log P_{Y|X}(y|x_t(m)) \nonumber \\
	&&+\frac{ 1}{n_i} \log \Delta_{\mathcal{C},m},\\
	& =& -   \sum_{x\in\mathcal{X}} \sum_{y\in\mathcal{Y}} \E\left[\frac{1}{n_i}    \sum_{t=1}^{n_i}  \mathbbm{1}\left\{x_t(m)=x,{Y}_{\mathcal{C},t}=y \right \} \Big|\tilde{M}=m\right] \nonumber \\
	&& \hspace{2cm}\cdot\log P_{Y|X}(y|x)  \nonumber \\
	&& +\frac{ 1}{n_i} \log \Delta_{\mathcal{C},m}, \\[1.2ex]
	& =& - \sum_{x\in\mathcal{X}} \sum_{y\in\mathcal{Y}} \sum_{s\in\mathcal{S}}  \E\left[ \pi_{{x}^{n_i}(m)S^{n_i}_{\mathcal{C}} {Y}_{\mathcal{C}}^{n_i}}(x,s,y)\Big|\tilde{M}=m\right]\nonumber \\[1.2ex]
	&& \hspace{2cm}\cdot\log P_{Y|X}(y|x)  \nonumber \\
	&&+\frac{ 1}{n_i} \log \Delta_{\mathcal{C},m},
\end{IEEEeqnarray}
where $P_{Y|X}(y|x)=\sum_{s\in\mathcal{S}} P_{Y|XS}(y|x,s) P_S(s)$.
Averaging over all messages $m\in \tilde{\mathcal{M}}$, we obtain:
\begin{IEEEeqnarray}{rCl}
	\lefteqn{\frac{1}{n_i}H({Y}^{n_i}_{\mathcal{C}} |\tilde{M}) }\\
	&\geq & - \sum_{x\in\mathcal{X}}   \sum_{y\in\mathcal{Y}}  \sum_{s\in\mathcal{S}}   \E\left[ \pi_{{x}^{n_i}(\tilde{M})S^{n_i}_{\mathcal{C}}{Y}_{\mathcal{C}}^{n_i}}(x,s,y)  \right] \cdot \log P_{Y|X}(y|x) \nonumber \\
	&& +   \E\left[ \frac{ 1}{n_i} \log \Delta_{\mathcal{C},\tilde{M}}\right].
\end{IEEEeqnarray}

By \eqref{eq:conditions} the term  $E\left[ \frac{ 1}{n_i} \log \Delta_{\mathcal{C},\tilde{M}}\right]$ vanishes for increasing blocklengths, and thus using the definition of  $P_X$ in \eqref{eq:convergence_point}, 
fone can follow the same bounding steps as leading to \eqref{eq:intermediate} to obtain: 
\begin{IEEEeqnarray}{rCl}
	\lefteqn{\lim_{i\to\infty} \frac{1}{n_i} H(\tilde{Y}^{n_i} |\tilde{M}) } \nonumber \\
	&= &- \sum_{x\in\mathcal{X}} {P}_X(x) \sum_{y \in \mathcal{Y}} \sum_{s\in\mathcal{S}} P_S(s) P_{Y|XS}(y|x,s) \log P_{Y|X}(y|x) \nonumber \\
	&=&H_{P_{X}P_SP_{Y|XS}}(Y|X).\label{eq:cond}
\end{IEEEeqnarray}

Combining \eqref{eq:diff} with \eqref{eq:limitHy} and \eqref{eq:cond}, and since $\frac{1}{n_i} \log \gamma \to 0$  as $n\to \infty$, we can conclude that 
\begin{IEEEeqnarray}{rCl}
	R &\leq & H_{P_{X}P_SP_{Y|XS}}(Y) - H_{P_{X}P_SP_{Y|XS}}(Y|X) \\
	&=&  I_{P_{X}P_SP_{Y|XS}}(X;Y).\label{eq:last}
\end{IEEEeqnarray}

\vspace{3mm}

\textbf{Proof of Distortion Bound:}
Consider the two conditions
\begin{subequations}\label{eq:sub2}
	\begin{IEEEeqnarray}{rCl} 
		\textnormal{dist}^{(n)}\left(h^{(n)}\left(x^n(m),z^n\right) , s^n\right) & \leq & D \nonumber\\ \label{eq:cond2} \\
		\left| \pi_{s^n,x^n(m),z^n}(a,b,c) - P_S(a) \pi_{x^n(m)}(b) P_{Z|XS}(c|a,b) \right| &\leq & \mu_n \label{eq:cond3b}, \nonumber\\
	\end{IEEEeqnarray}
\end{subequations}
and define for  each message $m \in \tilde{\mathcal{M}}$ the set 
\begin{IEEEeqnarray}{rCl} 
	\mathcal{D}_{\mathcal{S},m}& :=&  \left\{  (s^n,z^n)  \colon   \quad  \eqref{eq:cond2} \; \textnormal{and} \; \eqref{eq:cond3b} \right\}. \IEEEeqnarraynumspace 
\end{IEEEeqnarray}

Recall the definition $\tilde{X}^n=x^n(\tilde{M})$ and the limit in \eqref{eq:convergence_point}.  
Define  the   new random variables $({S}^n_{\mathcal{S}}, {Z}^n_{\mathcal{S}})$ of  joint conditional pmf
\begin{IEEEeqnarray}{rCl} 
	\lefteqn{
		P_{S^n_{\mathcal{S}}{Z}^{n}_{\mathcal{S}}|\tilde{M}} (s^n,z^n|m) } \nonumber \\
	&=&\frac{ P_S^{\otimes n}(s^n) \cdot P_{Z|XS}^{\otimes n}( z^n |x^n(m), s^n)}{  \Delta_{\mathcal{S},m}}  \cdot \mathbbm{1} \left\{ (s^n,z^n) \in  \mathcal{D}_{\mathcal{S},m} \right\}, \nonumber\\
\end{IEEEeqnarray}
for 
\begin{IEEEeqnarray}{rCl}\label{eq:Delta}
	\lefteqn{\Delta_{\mathcal{S},m} := \sum_{s^n,z^n} P_S^{\otimes n}(s^n) \cdot P_{Z|XS}^{\otimes n}(z^n |x^n(m), s^n)  } \hspace{4cm} \nonumber\\
	& & \cdot   \mathbbm{1} \left\{ (s^n,z^n) \in  \mathcal{D}_{\mathcal{S},m} \right\}.\IEEEeqnarraynumspace
\end{IEEEeqnarray}
Notice that by  \cite[Remark to Lemma~2.12]{Csiszarbook} and Conditions \eqref{eq:sub_exp2} and  \eqref{eq:sub2}, we have:
\begin{IEEEeqnarray}{rCl}
	\Delta_{\mathcal{S},m}\geq \eta-\frac{|\mathcal{S}||\mathcal{X}||\mathcal{Z}|}{4\mu_n^2 n}, \qquad \forall m \in \tilde{\mathcal{M}}. \IEEEeqnarraynumspace
\end{IEEEeqnarray}

Following similar steps to \eqref{eq:one}--\eqref{eq:PYlimit}, by \eqref{eq:cond3b} and definition \eqref{eq:convergence_point},  we can conclude that 
\begin{IEEEeqnarray}{rCl}
	\lim_{n_{i} \to \infty} P_{ \tilde{X}_T{S}_{\mathcal{S},T}{Z}_{\mathcal{S},T}}(x,s,z) = {P}_X(x) P_{S}(s)P_{Z|XS}(z|x,s). \nonumber \\
	\label{eq:PYlimit2}
\end{IEEEeqnarray}

By Condition~\eqref{eq:cond2}, we have  with probability 1: 
\begin{IEEEeqnarray}{rCl}
	D & \geq &\frac{1}{n}  \sum_{t=1}^{n} d \left( \hat{s}\left(\tilde{X}_t, Z_{\mathcal{S},t}\right) \;, \; S_{\mathcal{S},t}\right).
\end{IEEEeqnarray}
Therefore, 
for any blocklength $n_{i}$:
\begin{IEEEeqnarray}{rCl}
	D & \geq &\frac{1}{n_i}  \sum_{j=1}^{n_i}  \mathbb{E}\left[ d \left( \hat{s}\left(\tilde{X}_j, Z_{\mathcal{S},j}\right) \;, \; S_{\mathcal{S},j}\right)\right]  \\[1.2ex]
	& = & \mathbb{E}\left[ d \left( \hat{s}\left(\tilde{X}_T, Z_{\mathcal{S},T}\right) \;, \; S_{\mathcal{S},T}\right) \right],
\end{IEEEeqnarray}
and by \eqref{eq:PYlimit2} in the limit as $n_i\to \infty$:
\begin{IEEEeqnarray}{rCl}
	D & \geq & \mathbb{E}_{P_XP_SP_{Z|XS}}\left[ d \left( \hat{s}\left(X, S\right) \;,  \; S\right) \right].
\end{IEEEeqnarray}
This concludes the proof of the converse.

\section{Stein's Exponent as a  Sensing Measure}\label{sec:distortion}
In this section we assume that the state-sequence $S^n$ depends on a binary hypothesis $\mathcal{H}\in\{0,1\}$. Under the null hypothesis $\mathcal{H}=0$ it is i.i.d. according to the pmf $P_S$ and under the alternative hypothesis $\mathcal{H}=1$ it is  i.i.d. according to the pmf $Q_S$. The radar receiver  attempts to guess the underlying hypothesis based on the inputs and backscattered signals, so it produces a guess of the form \begin{equation}
\hat{\mathcal{H}} = h^{(n)}(X^n,Z^n) \in\{0,1\}.
\end{equation}

Radar sensing performance is measured in terms of Stein's exponent. That means, it is required that the type-I error probability 
\begin{equation}
\alpha_{n}:=\Pr\left[  \hat{\mathcal{H}}=1|\mathcal{H}=0\right] 
\end{equation} stays below a given threshold, while the type-II error probability 
\begin{equation}
\beta_n :=\Pr\left[  \hat{\mathcal{H}}=0|\mathcal{H}=1\right] 
\end{equation}
should decay exponentially fast to 0 with largest possible exponent. 
\medskip


\begin{definition}
A rate-exponent pair $(R,E)$ is  $(\epsilon,\delta)$-achievable over the state-dependent  DMC $(\mathcal{X},\mathcal{Y}, P_{Y|XS})$ with state-distribution $P_S$, if there exists a sequence of encoding, decoding, and estimation functions $\{(\phi^{(n)}, g^{(n)}, h^{(n)})\}$ such that for each blocklength $n$ the average  probability of error satisfies 
\begin{equation}\label{eq:error_epsl}
\varlimsup_{n\to \infty} p^{(n)}(\textnormal{error}) \leq  \epsilon, \qquad \mathcal{H}\in\{0,1\},
\end{equation}
while the detection error probabilities satisfy:  
\begin{equation}\label{eq:T1}
\varlimsup_{n\to \infty}  \alpha_{n} \leq \delta,
\end{equation}
and 
\begin{equation}\label{eq:T2}
-\varliminf_{n\to \infty}  \frac{1}{n} \log \beta_{n} \geq E.
\end{equation}
\end{definition}

\begin{theorem}
For any $\epsilon, \delta \geq 0$ satisfying $\epsilon+\delta <1$, a rate-exponent pair $(R,E)$ is $(\epsilon,\delta)$-achievable, if and only if, there exists a pmf $P_X$ 
 satisfying
 \begin{equation}
R \leq  \min\{ I_{P_{X}P_SP_{Y|XS}}(X;Y),   I_{P_{X}Q_SP_{Y|XS}}(X;Y)\},
\end{equation}
and 
 \begin{equation}
E \leq   \mathbb{E}_{P_{X}}\left[D( P_{Z|X} \| Q_{Z|X}) \right] 
\end{equation}
where $P_{Z|X}$ and $Q_{Z|X}$ denote the conditional marginals of $P_{S}P_{Z|XS}$ and  $Q_{S}P_{Z|XS}$, respectively.

\end{theorem} 
\begin{IEEEproof}
Achievability follows by standard random coding for a compound channel and by applying a Neyman-Pearson test at the radar receiver. The converse is proved in Appendix~\ref{app:proof}.\end{IEEEproof}

\medskip

The works in \cite{Joudeh2022Discrim,Bloch2022}  consider degenerate state-distributions where $P_S$ and $Q_S$ are deterministic distributions. In this case, our Theorem simplifies as follows.\footnote{Recall that  \cite{Joudeh2022Discrim,Bloch2022,Joudeh2021JBinaryDetect}  required exponential decrease both for the type-I and type-II error probabilities $\alpha_n$ and $\beta_n$. Their result is thus not a special case of ours.} 

\begin{corollary}
Assume degenerate state-distributions $P_{S}(s_0)=1$ and $Q_S(s_1)=1$ for two distinct symbols $s_0,s_1\in\mathcal{S}$. Then, for any $\epsilon, \delta \geq 0$ satisfying $\epsilon+\delta <1$, a rate-exponent pair $(R,E)$ is $(\epsilon,\delta)$-achievable, if and only if, there exists a pmf $P_X$ 
 satisfying
 \begin{equation}
R \leq  \min\left\{ I_{P_{X}P_{Y|X}^{(s_0)}}(X;Y), \;  I_{P_{X}P_{Y|X}^{(s_1)}}(X;Y)\right\},
\end{equation}
and 
 \begin{equation}
E \leq   \mathbb{E}_{P_{X}}\left[D( P_{Z|XS}(\cdot|X,s_0)\| P_{Z|XS}(\cdot|X,s_1) \right] ,
\end{equation}
where $P_{Y|X}^{(s)}(y|x)\triangleq P_{Y|XS}(y| x,s)$ for any triple $(x, s, y)$.
\end{corollary}

\section{Conclusion and Future Directions}
In this paper we established the strong converse for two ISAC problems with bi-static radar whenever $\epsilon+\delta < 1$. Interesting future research directions include extensions to mono-static radar systems where the transmitter can apply closed-loop encodings depending also on past generalized feedback systems or  systems with memory. Analyzing other sensing criteria is also of interest, such as the minimum exponential decay-rate over all hypotheses or the estimation error when the distribution of the state-sequence depends on a single continuous-valued parameter.
The setup where only part of the state or a noisy version of the state is to be estimated is also of interest, for instance, in scenarios in  state-dependent fading channels where one 
	has no interest in estimating the fading. Notice that this setup is included in the rate-distortion model through an appropriate definition of the distortion measure. On a related note, our model also includes as special case the setups where the receiver has perfect or imperfect channel-state information by including this state-information as part of the output. 

\clearpage
\bibliographystyle{ieeetran}
\bibliography{main.bib}

\clearpage

\appendices 
\section{Strong Converse Proof}\label{app:proof}

Fix a sequence of  encoding,  decoding, and estimation functions $\{(\phi^{(n)}, g^{(n)}, h^{(n)})\}_{n=1}^\infty$. Assume that   \eqref{eq:error_epsl}  and \eqref{eq:T1} are satisfied. For readability, we will also write $x^n(\cdot)$ for the function $\phi^{(n)}(\cdot)$. 
Choose a sequence of small positive numbers  $\{\mu_n \}$ satisfying
\begin{eqnarray}
\lim_{n\to \infty} \mu_n  & =& 0 \label{eq:mun1dl}\\
\lim_{n\to \infty} \left(n \cdot \mu_n^2  \right)^{-1}& =& 0. \label{eq:mun2dl}
\end{eqnarray}

\textbf{Expurgation:} Fix $\eta \in (0,1-\epsilon-\delta]$ and let  $\tilde{\mathcal{M}}$ be the set of messages $m$ that satisfy the following two conditions:
\begin{subequations}\label{eq:sub_expl}
\begin{IEEEeqnarray}{rCl} 
\Pr\left[\hat{M} \neq M|M=m \right] &\leq&1-\eta \IEEEeqnarraynumspace \label{eq:sub_exp1l}
\\
\Pr\left[  \hat{\mathcal{H}}=1\Big|\mathcal{H}=0 , M=m \right] &\leq& 1-\eta .\label{eq:sub_exp2l}
 \end{IEEEeqnarray}
 \end{subequations}
 The size of the set  $\tilde{\mathcal{M}}$ satisfies 
\begin{equation}
	\frac{|\tilde{\mathcal{M}}|}{2^{nR}} \geq \left(1-  \frac{\epsilon+\delta}{1-\eta}\right)= : \gamma.\end{equation}
 Define the random variable $\tilde{M}$ to be uniform over the set $\tilde{\mathcal{M}}$ and $\tilde{X}^n=x^n(\tilde{M})$.
 
 Let $T$ be uniform over $\{1,\ldots, n\}$ independent of all other quantities  and notice that 
\begin{IEEEeqnarray}{rCl} 
P_{\tilde{X}_T}(x)&= &  \mathbb{E}[\pi_{{x}^{n}(\tilde{M})}(x) ]. 
 \end{IEEEeqnarray}

Consider an increasing subsequence of blocklengths $n_i$ so that the probabilities $P_{\tilde{X}_T}(x)$ converge and denote the convergence point by $P_X$: 
 \begin{equation}\label{eq:limitXl}
\lim_{n_i \to \infty} P_{\tilde{X}_{T} }= P_{X}(x), \qquad x \in \mathcal{X}.
\end{equation} 
 \medskip

\textbf{Proof of Channel Coding Bound: } Considering the sequence of blocklengths $\{n_i\}_{i=1}^\infty$ and following  steps \eqref{eq:sub}--\eqref{eq:last} in the previous Section~\ref{sec:converseD}, once with the pmf $P_S$ and  once with the pmf $Q_S$, one can show that 
\begin{equation}
R \leq \min\left\{ I_{P_XP_{S}P_{Y|XS}}(X;Y), \;  I_{P_XQ_{S}P_{Y|XS}}(X;Y)\right\},
\end{equation}
where $P_X$ denotes the pmf in \eqref{eq:limitXl}. 

\medskip

\textbf{Proof of Stein's Exponent:}
Consider conditions 
\begin{subequations}\label{eq:sss}
\begin{IEEEeqnarray}{rCl} 
h^{(n)}\left(x^n(m),z^n\right)  ) & = & 0\nonumber \\ \label{eq:cond2l} \\
 \left| \pi_{s^n,x^n(m),z^n}(a,b,c) - P_S(a) \pi_{x^n(m)}(b) P_{Z|XS}(c|a,b) \right| &\leq & \mu_n \label{eq:cond3bl}, \nonumber\\
 \end{IEEEeqnarray}
 \end{subequations}
and define for  each message $m \in \tilde{\mathcal{M}}$ the set 
\begin{IEEEeqnarray}{rCl} 
\mathcal{D}_{\mathcal{S},m}& :=&  \left\{  (s^n,z^n)  \colon   \quad  \eqref{eq:cond2l} \; \textnormal{and} \; \eqref{eq:cond3bl} \right\}. \IEEEeqnarraynumspace 
\end{IEEEeqnarray}
Define  the   new random variables $({Y}^n_{\mathcal{S}}, {Z}^n_{\mathcal{S}})$ of  joint conditional pmf
\begin{IEEEeqnarray}{rCl} 
\lefteqn{
P_{S^n_{\mathcal{S}}  {Z}^{n}_{\mathcal{S}}|\tilde{M}} (s^n, z^n|m) } \nonumber \\
&=&\frac{ P_S^{\otimes n}(s^n) \cdot P_{Z|XS}^{\otimes n}( z^n |x^n(m), s^n)}{  \Delta_{\mathcal{S},m}}  \cdot \mathbbm{1} \left\{ (s^n,z^n) \in  \mathcal{D}_{\mathcal{S},m} \right\},\nonumber\\
\end{IEEEeqnarray}
for 
\begin{IEEEeqnarray}{rCl}\label{eq:Deltal}
\lefteqn{\Delta_{\mathcal{S},m} := \sum_{s^n,z^n} P_S^{\otimes n}(s^n) \cdot P_{Z|XS}^{\otimes n}( z^n |x^n(m), s^n)  } \hspace{4cm} \nonumber\\
& & \cdot   \mathbbm{1} \left\{ (s^n,z^n) \in  \mathcal{D}_{\mathcal{S},m} \right\}.\IEEEeqnarraynumspace
\end{IEEEeqnarray}
By  \cite[Remark to Lemma~2.12]{Csiszarbook} and Conditions \eqref{eq:sub_exp2l} and  \eqref{eq:sss}, we have:
\begin{IEEEeqnarray}{rCl}
\Delta_{\mathcal{S},m}\geq \eta -\frac{|\mathcal{S}||\mathcal{X}||\mathcal{Z}|}{4\mu_n^2 n}, \qquad \forall m \in \tilde{\mathcal{M}}. \label{eq:DSM}\IEEEeqnarraynumspace
\end{IEEEeqnarray}

Following similar steps to \eqref{eq:one}--\eqref{eq:PYlimit}, we can conclude that 
\begin{IEEEeqnarray}{rCl}
\lefteqn{\lim_{n_{i} \to \infty} P_{ \tilde{X}_T{S}_{\mathcal{S},T}{Z}_{\mathcal{S},T}}(x,s,z)} \nonumber \qquad \\
 & = & {P}_X(x) P_{S}(s)P_{Z|XS}(z|x,s). \label{eq:PYlimit2l}
\end{IEEEeqnarray}

We next notice the inequalities:
\begin{IEEEeqnarray}{rCl}
\beta_{n} & = & \Pr[ \hat{\mathcal{H}}=0 |\mathcal{H}=1] \\
&= &  \frac{1}{2^{nR}} \sum_{m=1}^{2^{nR}} \Pr\left[ \hat{\mathcal{H}}=0 \Big|\mathcal{H}=1, M=m\right] \\
&\geq & \frac{|\tilde{\mathcal{M}}|}{2^{nR}} \cdot  \frac{1}{|\tilde{\mathcal{M}}|} \sum_{m\in\tilde{\mathcal{M}}} \Pr\left[ \hat{\mathcal{H}}=0 \Big |\mathcal{H}=1, \tilde{M}=m\right] \\
& \geq &   \gamma  \cdot \mathbb{E}_{\tilde{M}}\left[ \Pr\left[ \hat{\mathcal{H}}=0 |\mathcal{H}=1, \tilde{M}=m\right] \right],
\end{IEEEeqnarray}
and therefore, 
\begin{IEEEeqnarray}{rCl}
\lefteqn{- \frac{1}{n} \log \beta_n } \nonumber\\
& \leq & - \frac{1}{n} \log  \mathbb{E}_{\tilde{M}}\left[ \Pr\left[ \hat{\mathcal{H}}=0 \Big|\mathcal{H}=1, \tilde{M}=m\right] \right] - \frac{1}{n} \log   \gamma,\nonumber\\\label{eq:beta1}
\end{IEEEeqnarray}
where notice that the term $\frac{1}{n} \log   \gamma $ vanishes asymptotically for infinite blocklengths. 

Defining 
\begin{IEEEeqnarray}{rCl} Q_{Z|X}(z|x)&:= &\sum_{s} Q_S(s) P_{Z|SX}(z|x,w)\\
P_{Z|X}(z|x)&:= &\sum_{s} P_S(s) P_{Z|SX}(z|x,w),
\end{IEEEeqnarray}
 we can further obtain: 
\begin{IEEEeqnarray}{rCl}
\lefteqn{ - \frac{1}{n} \log  \mathbb{E}_{\tilde{M}}\left[ \Pr\left[ \hat{\mathcal{H}}=0 \Big|\mathcal{H}=1, \tilde{M}=m\right] \right]}\nonumber\\
& =& - \frac{1}{n} \log  P_{\tilde{X}^n} Q_{Z|X}^{\otimes n}(\hat{\mathcal{H}}=0) \\
& = &\frac{1}{n} D\left( P_{\tilde{X}^nZ_{\mathcal{S}}^n}  (\hat{\mathcal{H}}) \  \big\| \ P_{\tilde{X}^n} Q_{Z|X}^{\otimes n} (\hat{\mathcal{H}}) \right) \\
& \leq &\frac{1}{n} D\left( P_{\tilde{X}^nZ_{\mathcal{S}}^n} \  \big\| \ P_{\tilde{X}^n} Q_{Z|X}^{\otimes n} \right) \\
& = &\frac{1}{n} \sum_{m \in \tilde{\mathcal{M}}}  \frac{1}{|\tilde{\mathcal{M}}|} \sum_{z^n} P_{ Z_{\mathcal{S}}^n|\tilde{X}^n}(z^n|x^n(m)) \nonumber \\
&& \hspace{3cm} \cdot  \log \frac{ P_{ Z_{\mathcal{S}}^n|\tilde{X}^n}(z^n|x^n(m))}{ Q_{Z|X}^{\otimes n}(z^n|x^n(m) )}  \\
 & \leq &\frac{1}{n} \sum_{m \in \tilde{\mathcal{M}}}  \frac{1}{|\tilde{\mathcal{M}}|} \sum_{z^n} P_{ Z_{\mathcal{S}}^n|\tilde{X}^n}(z^n|x^n(m)) \qquad  \nonumber \\
&& \hspace{3cm} \cdot  \log \frac{ P_{ Z|X}^{\otimes n} (z^n|x^n(m))}{ Q_{Z|X}^{\otimes n}(z^n|x^n(m) )}\qquad   \nonumber\\
 && - \frac{1}{n} \sum_{m \in \tilde{\mathcal{M}}}  \frac{1}{|\tilde{\mathcal{M}}|}  \log \Delta_{\mathcal{S},m}\\
  & = &\frac{1}{n} \sum_{x^n}P_{\tilde{X}^n Z_{\mathcal{S}}^n}(x^n,z^n) \log \frac{ P_{ Z|X}^{\otimes n} (z^n|x^n)}{ Q_{Z|X}^{\otimes n}(z^n|x^n )} \nonumber \\
  && - \frac{1}{n} \mathbb{E}\left[  \log \Delta_{\mathcal{S},\tilde{M}} \right] \\
   & = &\frac{1}{n} \sum_{t=1}^n \sum_{x_t,z_t} P_{\tilde{X}_t Z_{\mathcal{S},t}} (x_t,z_t) \log \frac{ P_{ Z|X} (z_t|x_t)}{ Q_{Z|X}(z_t|x_t )}  \nonumber \\
   &&- \frac{1}{n} \mathbb{E}\left[  \log \Delta_{\mathcal{S},\tilde{M}} \right] \\
    & = & \sum_{x,z} P_{\tilde{X}_T Z_{\mathcal{S},T}} (x,z) \log \frac{ P_{ Z|X} (z|x)}{ Q_{Z|X}(z|x)}  \nonumber \\
    && - \frac{1}{n} \mathbb{E}\left[  \log \Delta_{\mathcal{S},\tilde{M}} \right].\label{eq:beta2}
\end{IEEEeqnarray}

Combining \eqref{eq:PYlimit2l}, \eqref{eq:beta1}, and \eqref{eq:beta2}, and considering the subsequence of blocklengths $\{n_i\}$, we conclude  that 
\begin{equation}
\varlimsup_{i\to \infty} - \frac{1}{n} \log \beta_{n_i} \leq \mathbb{E}_{P_X}\left[ D(P_{Z|X} \| Q_{Z|X})  \right],
\end{equation}
where we used that $\frac{1}{n} \mathbb{E}\left[  \log \Delta_{\mathcal{S},\tilde{M}} \right] \to 0$ as $n_i\to \infty$ by \eqref{eq:DSM}.
This concludes the proof of the converse.

\end{document}